# Observational distinguishability of the Kerr and Kerr-Hayward metrics to EHT

Nikola Bukowiecka [*]
*Department of Physics, University of Rhode Island,*
*2 Lippitt Road, Kingston, Rhode Island 02881-0817, USA*

Angelo Ricarte
*Center for Astrophysics—Harvard & Smithsonian, 60 Garden Street, Cambridge, Massachusetts 02138, USA*
*and Black Hole Initiative at Harvard University, 20 Garden Street, Cambridge, Massachusetts 02138, USA*

Prashant Kocherlakota
*Chennai Mathematical Institute,*
*India, H1, SIPCOT IT Park, Kelambakkam, Siruseri, Tamil Nadu 603103, India,*
*and Black Hole Initiative at Harvard University, 20 Garden Street, Cambridge, Massachusetts 02138, USA*

Cora Prather
*Black Hole Initiative at Harvard University, 20 Garden Street, Cambridge, Massachusetts 02138, USA*

Astrophysical black holes appear well-represented by the Kerr metric, whose maximal analytic extension contains a ringlike curvature singularity. We study a phenomenological regularized modification of Kerr, the modified Kerr-Hayward metric, which removes curvature blowup at the level of the stationary metric while preserving in detail many features of polarized black hole images now testable by the Event Horizon Telescope (EHT). To establish this, we produce new general relativistic magnetohydrodynamics (GRMHD) simulations of a magnetized plasma in a Kerr-Hayward spacetime, then we extend the EHT analysis framework to perform polarized radiative transfer in this spacetime. We detail our methodology for implementing this modified spacetime into an open-source pipeline. From fluid quantities such as the magnetic flux parameter and jet efficiency, to image quantities such as the polarization pattern and the photon ring structure, our results for the Kerr-Hayward metric appear functionally indistinguishable from the Kerr metric. Our study finds that under the assumptions of a fixed-background GRMHD and polarized radiative transfer pipeline, this regularized phenomenological modification to the Kerr metric can yield observables that are effectively indistinguishable in EHT measurements.

## I. INTRODUCTION

Using a worldwide network of millimeter telescopes, the Event Horizon Telescope (EHT) Collaboration is now producing images of black holes (BHs) on event horizon scales [1,2]. These images offer a way of probing general relativity (GR), our prevailing theory of classical gravity, in the strong-field regime [3–7]. At current resolution, these images reveal bright emission rings consistent with those in simulated Kerr black hole images [5,8].

Interpreting such images as tests of the spacetime metric, however, requires care: the observed emission depends not only on the background geometry, but also on the accretion flow, electron thermodynamics, magnetic-field structure, radiative transfer, and the image-reconstruction procedure. This point has been emphasized in previous work, which showed that horizon-scale images and ring features can be strongly affected by nongravitational plasma physics and therefore need not provide clean, model-independent tests of GR at current EHT resolution [9–13]. A complementary approach is to quantify these degeneracies directly by performing general relativistic magnetohydrodynamics (GRMHD) and general relativistic ray-tracing (GRRT) calculations in non-Kerr or phenomenological spacetimes and comparing the resulting observables with Kerr models. Current EHT analyses also address astrophysical uncertainty through calibration over libraries of simulations; for example, the Sgr A* metric test used an image-to-shadow calibration factor to relate the observed ring diameter to the geometrically defined shadow size while marginalizing over model uncertainty [5,14–16]. Thus, EHT observations can provide meaningful spacetime constraints, but the strength and interpretation of those constraints depend on the astrophysical and emission-model assumptions.

[*]Contact author: nikola.bukowiecka@uri.edu

In the future, increases in dynamic range and spatial resolution will be made possible by extensions to the array both on the ground [17] and in space [18]. These may improve access to observables whose interpretation is expected to be more directly tied to the spacetime geometry, including the inner shadow (the direct image of the equatorial event horizon, relevant for disklike emission geometries [7]) and the photon ring (a sharp feature comprised of photons on marginally bound orbits around the BH [6]). This, however, has to go hand in hand with the development of more advanced analysis methods.

The Kerr metric [19] describes the spacetime of a stationary, rotating vacuum black hole in general relativity. Despite its central role in modeling astrophysical black holes, its interior geometry exhibits a key pathology. It contains a ringlike curvature singularity that leads to geodesic incompleteness. A complementary phenomenological approach has been introduced to construct nonsingular black hole models that remove the curvature singularity at the level of the stationary metric; see, e.g., Refs. [20–28]. These models (partly[1]) evade the Penrose singularity theorem [29] by relaxing its assumptions, typically the classical energy conditions, thereby avoiding curvature blowup in the stationary geometry. The resulting spacetimes are sourced by self-gravitating matter at the cost of introducing a physical problem of energy condition violation: the null energy condition is violated typically in the deep interior, within the inner horizon; see, e.g., Refs. [24,30]. The Raychaudhuri equation [31] then shows that such violations induce the defocusing of null congruences near the center, averting singularity formation. However, removing curvature blowup from the stationary metric does not by itself establish that the spacetime remains nonsingular under perturbations. In particular, the presence of an inner horizon raises the possibility of a mass-inflation instability, so the stability and geodesic completeness of such regular black hole geometries remain open issues; see, e.g., Refs. [32,33]; cf. also Ref. [34]. In spinning extensions, additional global pathologies can also arise, including regions with closed timelike curves [30]. These issues further motivate treating such metrics as phenomenological testbeds for horizon-scale observables rather than as complete physical spacetimes.

Among regular black hole spacetimes, the nonsingular model introduced in Ref. [23] serves as a particularly illustrative example. It is supported by matter with finite energy density and anisotropic pressures, and develops an effective de Sitter core, corresponding to a positive cosmological constant, in the vicinity of the regular center. This behavior has previously been proposed as the appropriate effective equation of state of matter at high densities, motivated by the assumption of an upper limit on density or curvature; see, e.g., Ref. [35].[2] Rotating generalizations of this BH model have also been constructed using the Newman-Janis trick [37] in [24].

In this work, we focus on the spinning generalization of a minimally modified Kerr-Hayward geometry, which we refer to henceforth as the modified Kerr-Hayward metric. Specifically, we consider the spinning generalization of the $n = 4$ Hayward model introduced in Ref. [27], following the description in Ref. [30]. We emphasize that the modified Kerr-Hayward metric is used here as a phenomenological spacetime rather than as a solution derived from a complete theory of gravity and matter. Consequently, our results should not be interpreted as predictions of a fully self-consistent alternative gravity model. A complete theory that produced this metric could also modify the dynamics of plasma and electromagnetic fields, for example through couplings to the stress energy sourcing the background geometry. In this work, we neglect such effects and instead ask a narrower question: whether, within the standard test-fluid GRMHD and GRRT framework used in EHT modeling, this representative regular black hole spacetime produces observables distinguishable from Kerr. The nonspinning case is not only free of curvature singularities, but is also geodesically complete in its maximal extension. Its spinning extension has not yet been analyzed for its geodesic completeness. Moreover, as for other multi-horizon black hole spacetimes, the inner horizon may be unstable to mass inflation under perturbations. We therefore use the modified Kerr-Hayward metric as a phenomenological testbed for exterior and horizon-scale observables rather than as a complete resolution of black hole singularities. These geometries are characterized by three parameters: the total mass, the total angular momentum, and the de Sitter length scale associated with the core mass compactness or, equivalently, the central cosmological constant. For such BH models, deviations of the spacetime metric from the Kerr solution are appreciable only in the immediate vicinity of the horizon. This makes the modified Kerr-Hayward metric a useful phenomenological testbed for assessing whether the EHT, or its future extensions, can observationally distinguish this class of regular black hole models from their Kerr counterparts.

[1] These spacetimes are curvature singularity free but typically still geodesically incomplete [27].

[2] This is conceptually analogous to the field-strength cutoff in Born-Infeld nonlinear electrodynamics, often viewed as an effective vacuum-polarization correction to Maxwell theory; see, e.g., Ref. [36].

In the context of gravity tests with EHT observations and horizon-scale imaging, the emerging state-of-the-art is to run new GRMHD simulations in exotic spacetimes [38–44]. Thus, in this work, we have extended the current EHT analysis framework to accommodate the modified Kerr-Hayward BH model: from 3D GRMHD simulations, to GRRT, to the calculation of polarized image characteristics for direct comparison with EHT observations. The purpose of this extension is to determine whether EHT and BHEX could distinguish between the Kerr and modified Kerr-Hayward metric. If we assume that certain properties of the Kerr metric generalize to the modified Kerr-Hayward metric,[3] then for M87* models, we identify no signatures that distinguish modified Kerr-Hayward from Kerr, even for large interior core sizes. We discuss the limitations of the used methodologies further in the conclusions. Our goal is not to establish a general limitation on EHT tests of GR, but to ask whether this particular phenomenological regular spacetime produces observable differences once the full GRMHD + GRRT modeling pipeline is applied. Conversely, other parameterized or theory-motivated non-Kerr spacetimes can produce larger image deviations, so the degeneracy found below should be understood as model-dependent rather than universal [41,42].

In Sec. II, we motivate and introduce the modified Kerr-Hayward spacetime and discuss the implementation details. All GRMHD simulation details, including necessary parameters to reproduce our runs, are listed in Sec. II B, while all polarized ray-tracing parameters can be found in Sec. II C. We present our results in Sec. III, divided into *Fluid domain* GRMHD quantities (accretion rate, magnetic flux, and jet efficiencies) and *Imaging and polarization domain* observables from GRRT (total intensity, polarization metrics, the inner shadow, and the photon ring). Finally, in Sec. IV, we discuss and summarize our results and their significance, along with the limitations of our methodologies.

[3]The symmetries and the Petrov classification of the modified Kerr-Hayward spacetime have been discussed in Ref. [45]. Like the Kerr metric, these are Type D spacetimes and have a Killing as well as a Killing-Yano tensor. We therefore expect that the Teukolsky equation for the modified Kerr-Hayward spacetime should be trivially and smoothly modified from its Kerr form. Further, the Quasinormal (QNM) mode spectrum for the modified Kerr-Hayward metric should shift slightly and smoothly from that for Kerr. If these expectations hold true, then the Kerr-Hayward BH is mode stable to QNMs, just as Kerr is.

## II. METHODS

### A. Modified Kerr-Hayward spacetime and implementation

The spherically symmetric (nonspinning) modified Hayward metric is described, in areal-polar coordinates, by the line element (see Eqs. 9 and 13 of Ref. [27] with $n = 4$)

$$ds^2 = -f(r)\mathrm{d}t^2 + \frac{\mathrm{d}r^2}{f(r)} + r^2\mathrm{d}\Omega_2^2,$$
$$\text{with} \quad f(r) = 1 - \frac{2Mr^3}{r^4 + 2L^4}\left(=: 1 - \frac{2m(r)}{r}\right). \qquad (1)$$

In the above, $\mathrm{d}\Omega_2^2$ denotes the standard line element on the unit 2-sphere, $M$ is the total (ADM) mass of the spacetime, and $m(r)$ is the Misner-Sharp/Hawking mass enclosed within a sphere of radius $r$. As noted earlier, both the Weyl and Kretschmann scalars remain finite everywhere, and the (Hayward) spacetime is geodesically complete in its maximal extension.

This metric describes BHs with two horizons when the de Sitter length scale $L$ satisfies $L < L_{\mathrm{ex}} = (27/32)^{1/4}$. At $L = L_{\mathrm{ex}}$, the horizons coincide, yielding an extremal Hayward black hole. For $L > L_{\mathrm{ex}}$, no horizons are present, and the spacetime instead describes a regular, horizonless matter distribution analogous to a compact star. The radius of the extremal horizon is $r_{\mathrm{H;ex}} = 1.5M$, which is 25% smaller than that of a Schwarzschild black hole with the same ADM mass.

The energy–momentum tensor $\mathcal{T}_{\mu\nu}$ of the self-gravitating matter can be obtained from the Einstein tensor $\mathcal{G}_{\mu\nu}$ via the field equations, $\mathcal{G}_{\mu\nu} = 8\pi\mathcal{T}_{\mu\nu}$. The corresponding energy density and pressure profiles are shown in Fig. 2 of Ref. [30]. The energy density $\epsilon$ is non-negative, while the normal (radial) pressure $p_n$ differs from the tangential pressures, indicating anisotropy. The radial equation of state is constant, $p_n/\epsilon = -1$, thereby marginally saturating the null energy condition. In contrast, the tangential equation of state is radius-dependent and violates the null energy condition inside the inner horizon. In the exterior region, the null energy condition is satisfied, and only the dominant energy condition is violated.

In line with previous approaches used to produce spinning counterparts to regular BH models, we employ the Newman-Janis-inspired Azreg-Aïnou algorithm [46] with the modified Hayward metric (1) as the input “seed” metric,[4] to obtain the modified Kerr-Hayward metric, in Boyer-Lindquist coordinates, as [30],

[4]See, however, [45] for a discussion on the physical viability and limits of such algorithms.

$$ds^2 = -\left(1 - \frac{2F}{\Sigma}\right)dt^2 - 2\frac{2F}{\Sigma}a\sin^2\vartheta\, dt d\varphi + \frac{\Pi}{\Sigma}\sin^2\vartheta\, d\varphi^2 + \frac{\Sigma}{\Delta}dr^2 + \Sigma d\vartheta^2, \quad (2)$$

where the spinning metric functions $F$, $\Delta$, $\Sigma$, and $\Pi$ are related to the nonspinning metric function $f$ in Eq. (1) via the algorithm as

$$\begin{aligned} 2F(r) &= (1 - f(r))r^2, \\ \Delta(r) &= f(r)r^2 + a^2, \\ \Sigma(r, \vartheta) &= r^2 + a^2\cos^2\vartheta, \\ \Pi(r, \vartheta) &= (r^2 + a^2)^2 - \Delta(r)a^2\sin^2\vartheta. \end{aligned} \quad (3)$$

In the limit $L \to 0$, the modified Kerr-Hayward metric reduces to Kerr. More generally, for any fixed $L$, the spacetime rapidly approaches Kerr with increasing distance from the horizon. At fixed $M$, the BH parameter space $(a, L)$ closely parallels that of the Kerr-Newman family. Increasing either $L$ or $a$ shifts a greater fraction of the mass-energy to the exterior, thereby reducing the horizon (Hawking) mass. Since this satisfies $m(r_\mathrm{H}) = \sqrt{\mathcal{A}_\mathrm{H}/(16\pi)}$, with $\mathcal{A}_\mathrm{H}$ the horizon area, the areal radius decreases monotonically toward extremality. In the extremal limit, the horizon size is reduced by $\sim$27% relative to a Schwarzschild black hole of the same ADM mass.

We present a brief overview of the properties of the self-gravitating matter in this spacetime (for further details, see Sec. 5.2 of Ref. [30]). On each Boyer-Lindquist coordinate sphere, the matter undergoes rigid rotation about the spin axis with angular velocity $\Omega(r) = a/(r^2 + a^2)$. The energy density is nonnegative everywhere, and the normal equation of state remains fixed, $p_n/\epsilon = -1$. In contrast to the spherically symmetric case, the tangential pressures $(p_\vartheta, p_\varphi)$ are unequal and exhibit nontrivial dependence on both radius and colatitude $\vartheta$. The null energy condition is violated in the deep interior (inside the inner horizon), while in the BH exterior all classical energy conditions, except the dominant energy condition, are satisfied.

Although the modified Kerr-Hayward metric is free of curvature singularities, its geodesic completeness has not yet been established. Moreover, these spinning spacetimes contain pathological regions admitting closed timelike curves in the Boyer-Lindquist $r < 0$ region (see Sec. 5.1 of Ref. [45]). These solutions belong to the class of "degenerate ACKN spacetimes" introduced in Ref. [45], and consequently admit separability of the geodesic and Klein-Gordon equations and likely also of the Dirac equation. They possess both a Killing tensor and a Killing-Yano tensor, and are algebraically special (Petrov type D) spacetimes.

Finally, to facilitate GRMHD simulations, we express the modified Kerr-Hayward metric (2) in a horizon-penetrating coordinate system, namely the "siKS coordinates" of Ref. [47], as

$$ds^2 = -\left(1 - \frac{2F}{\Sigma}\right)d\tau^2 + \left(1 + \frac{2F}{\Sigma}\right)dr^2 + \Sigma d\vartheta^2 + \frac{\Pi}{\Sigma}\sin^2\vartheta\, d\phi^2 - 2\frac{2F}{\Sigma}a\sin^2\vartheta\, d\tau d\phi + 2\frac{2F}{\Sigma}d\tau dr - 2\left(1 + \frac{2F}{\Sigma}\right)a\sin^2\vartheta\, dr d\phi. \quad (4)$$

In these coordinates, $x^{\bar{\mu}} = (\tau, r, \vartheta, \phi)$, the metric is regular at the horizon, enabling stable numerical evolution and the evaluation of fluid variables arbitrarily close to the horizon.

### B. GRMHD

We performed all of our numerical simulations using Kokkos-based High-Accuracy Relativistic Magnetohydrodynamics with Adaptive mesh refinement (KHARMA) which, among others, implements the High-Accuracy Relativistic Magnetohydrodynamics (HARM) scheme [48]. Due to the nature of the MHD equations, we are considering the "ideal-MHD" limit, where we assume the electrical conductivity to be infinite, which is a standard approximation in the EHT analysis. In the MHD approximation, we treat plasma as a macroscopic fluid coupled to the electromagnetic field, disregarding the microscopic kinetic effects. For familiarization with the conservative, shock-capturing scheme for evolving the GRMHD equations solved in KHARMA, we refer the reader to [49].

Throughout this work, the metric remains fixed and the GRMHD plasma is treated as a test fluid on that background. Therefore, the self-gravitating matter sourcing the modified Kerr-Hayward spacetime is not evolved, and no nongravitational interaction between that matter and the accreting plasma or electromagnetic field is included. This is the same approximation implicit in using a prescribed background spacetime in standard GRMHD simulations, but in the present phenomenological nonvacuum metric, it should be regarded as an explicit modeling assumption. If a complete theory generating the Kerr-Hayward metric introduced additional matter–sector couplings, the corresponding MHD equations could differ from those solved here. In the absence of such a theory, introducing additional couplings would be arbitrary, so we leave this possibility to future work.

We generalized the current scheme by implementing the modified Kerr-Hayward metric in KHARMA in the form given by Eq. (4), and we ran simulations for the varying dimensionless spin $a = 0.1$, 0.5 and the length scale parameter $L = 0$, 0.5. The case $L = 0$ recovers the Kerr metric, while $L = 0.5$ is chosen to introduce moderate, representative deviations from Kerr. The implementation and procedure documentation have been made open-source [50].[5] We define $t_g \equiv \frac{GM}{c^3}$, $r_g \equiv \frac{GM}{c^2}$ and set $G = c = 1$.

We initialized an equilibrium gas torus [51] parameterized by $r_{\rm in} = 20\ {\rm r_g}$ and $r_{\rm max} = 41\ {\rm r_g}$. We simulated our models for $t_{\rm lim} = 30000\ {\rm t_g}$, but after empirically confirming that all four simulations reach a quasisteady state, we consider the latter $t = 15000\ {\rm t_g}$. We adopt the Magnetically Arrested Disc (MAD) initial magnetic field configuration used in EHTC *et al.* [52], which threads the torus with a single poloidal field loop with initial $\beta = p_{\rm gas}/p_{\rm mag} = 100$. MAD refers to an accretion state where the magnetic flux accumulates near the black hole until it becomes dynamically important and partially arrests the inflow [53–55]. Our grid spans the radius from $1.36\ {\rm r_g} - 1000\ {\rm r_g}$ and full polar and azimuthal domains $(0 \le \theta \le \pi)$ and $(0 \le \phi \le 2\pi)$, respectively. The base mesh resolution was $96 \times 64 \times 64$ and the timestep was set to 0.7 of the CFL stability limit for our second-order scheme. Accretion in our highly magnetized systems is driven by differential rotation, not the magnetorotational instability, and is unaffected by low resolution. We comparatively examine snapshots and time-averaged state, which are less vulnerable to low resolution than the system variability. We used an adiabatic equation of state with an adiabatic index of $5/3$ [56].

We follow the PATOKA pipeline [57] to simulate accretion flows of four fiducial models, which we henceforth label as: *KerrLow* or a0.1L0 with $a = 0.1$, *KerrMid* or a0.5L0 with $a = 0.5$, *KHLow* or a0.1L0.5 with $a = 0.1$, $L = 0.5$, *KHMid* or a0.5L0.5 with $a = 0.5$, $L = 0.5$, where $a$ denotes the dimensionless spin. All scripts used in analyzing this data, along with the simulation logs, have been made open-source [50] and use PYHARM.[6]

### C. GRRT

The GRMHD simulations were post-processed using a polarized covariant radiative transfer code IPOLE [58], as described in [57]. Each polarimetric image contains information about the Stokes parameters for each pixel, where linear polarization fraction is defined as $LP = \sqrt{Q^2 + U^2}/I$, and the image defined by a field of view in μas, distance from the black hole, and inclination $i$.

Considering that GRMHD simulations allow for rescaling in which the BH mass sets the spacetime length and time units, while a separate normalization parameter, $M_{\rm unit}$, rescales the fluid density, internal energy, and magnetic field strength, provided that dimensionless plasma quantities (such as the plasma beta and magnetization) remain unchanged [59]. Here, instead of adopting a single constant normalization, following [59], we introduce a time-dependent scaling $M(t) = \exp(a + bt)$, which is fit to the simulation data. We perform this parameterization to match the observed flux density and compensate for a secular decline in accretion rate. The parameters $a$ and $b$ are determined by fitting to the GRMHD frames using a Nelder-Mead optimization procedure. The resulting $M(t)$ defines the effective mass accretion normalization $M_{\rm fit}$, which is then used as input to the GRRT calculations to generate synthetic images.

For our postprocessing, we have chosen the typical EHT observation parameters for $M87^*$: $M^* = 6.5 \times 10^9 M_\odot$, $D = 16.8$ Mpc, $i = 163°$. We compute images at $400 \times 400$ pixel resolution and a frequency of 230 GHz. The low polarization fraction of EHT images of M87* favors models where electrons are significantly cooler than ions, as expected in radiatively inefficient flows [60,61]. We adopt the $R - \beta$ prescription of Mościbrodzka *et al.* [62] (Eqs. 1) to set the electron temperature:

$$\frac{T_p}{T_e} = R_{\rm high}\frac{b^2}{1 + b^2} + R_{\rm low}\frac{b^2}{1 + b^2}, \qquad (5)$$

where $b = \beta/\beta_{\rm crit} = \beta/1$ and $R_{\rm high}, R_{\rm low}$ refer to asymptotic temperature ratios of electron-to-proton coupling in weakly (*high* $\beta$) and strongly (*low* $\beta$) magnetized regions. Here, we adopt $R_{\rm high} = 40, R_{\rm low} = 1$. We also zero the density in regions where plasma $\sigma \equiv b^2/\rho > 1$, where we expect contamination from floors.

To obtain the EHT observables, convolution of Stokes intensities is performed using a 20 μas Gaussian beam (the blurring kernel), to mimic present EHT resolution [1]. Once the blurring kernel is placed, from the Fourier decomposition of the complex linear polarization, the $\beta$-modes are calculated. Of particular interest to the EHT are the $\beta_2$ modes that quantify the rotationally symmetric polarization structure, so they inform the azimuthal (toroidal) polarization. This enables us to probe both emergent properties of the spacetime that may lead to differences in magnetic field structure as well as "gravitational Faraday rotation" [63].

We implement an analysis package for computing photon ring metrics (e.g., photon ring diameter and asymmetry) calculations, the comparison of Kerr and Kerr-Hayward metric, and we make it open-source: [50].

## III. RESULTS

### A. Fluid domain

The top panel in Fig. 1 shows the time evolution of the dimensionless Eddington ratio, calculated as [64]

$$M_{\rm Edd} = \frac{\dot{M}}{\dot{M}_{\rm Edd}}, \qquad (6)$$

where

[5]The KHARMA branch containing the extension to modified Kerr-Hayward spacetime (and illustrating the procedure of extending to arbitrary spacetime) can be found at https://github.com/nikolabukowiecka/kharma/tree/feature/KH-spacetime.

[6]https://github.com/AFD-Illinois/pyharm.

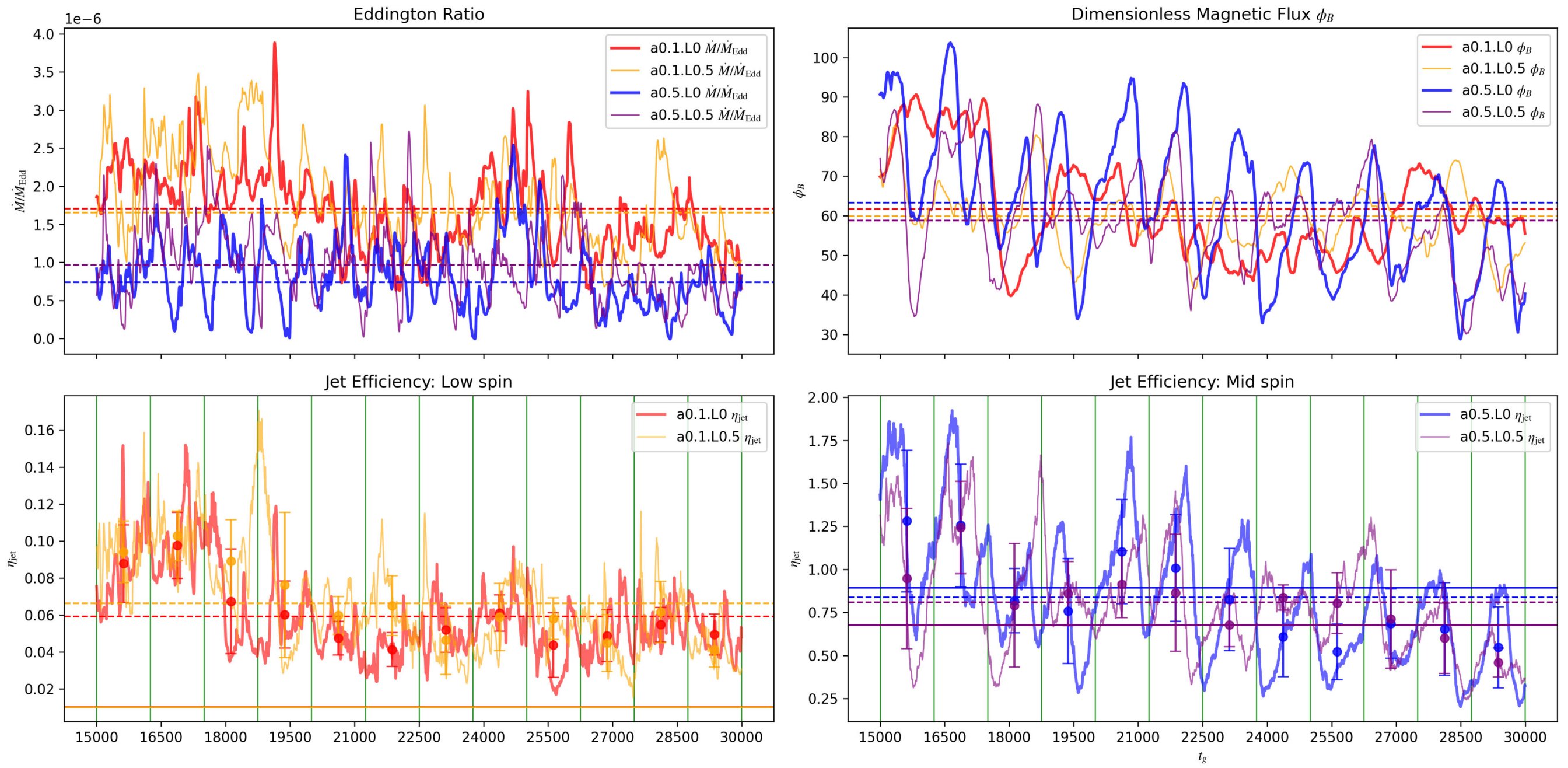

FIG. 1. Time evolution of fluid-domain observables for the four fiducial models. Top left: Eddington ratio, $\dot{M}/\dot{M}_{\rm Edd}$. Top right: dimensionless magnetic flux, $\phi_B$. In both top panels, the horizontal dashed lines denote the time-averaged values for each model. Bottom left: jet efficiency for the low-spin models. Bottom right: jet efficiency for the mid-spin models, where $\eta_{\rm jet}$ is defined in Eq. (10). In the bottom panels, the horizontal dashed lines mark the mean jet efficiencies, while the solid horizontal lines show the corresponding Blandford-Znajek estimates, $\eta_{\rm BZ}$, computed following Tchekhovskoy *et al.* [65]. The green vertical lines indicate the time bins used to compute the binned averages, shown with error bars corresponding to the standard deviation of the mean.

$$\dot{M}_{\rm Edd} = \frac{L_{\rm Edd}}{\epsilon c^2}, \tag{7}$$

and

$$L_{\rm Edd} = \frac{4\pi G M m_p c}{\sigma_T}. \tag{8}$$

Here, we assume a radiative efficiency $\epsilon = 0.1$, consistent with IPOLE. Both $\dot{M}$ and $\dot{M}_{\rm Edd}$ were read from IPOLE image headers and are thus scaled to the mass and specific luminosity of M87*. The values for low-spin models are higher than for the mid-spin ones, implying that they require slightly higher densities to reproduce the observed 0.5 Jy flux density of M87*. All models exhibit similar levels of variability. In the second panel of Fig. 1, we plot the dimensionless magnetic flux $\phi_B$ evolution, calculated as

$$\phi_B = \sqrt{4\pi}\frac{\Phi_B}{\sqrt{\langle \dot{M} \rangle}}, \tag{9}$$

where the factor of $\sqrt{4\pi}$ converts from Heaviside-Lorentz to Gaussian units and $\dot{M}$ is calculated at $r = 5M$ using PYHARM.

All four fiducial models achieve comparable MAD levels of $\phi_B \sim 60$. This is larger than the value of $\approx$50 reported by Narayan *et al.* [64] for these spin values, but the difference may be attributable to the difference in adiabatic index (13/9 in their study). We notice that $\phi_B$ is systematically lower in the modified Kerr-Hayward models compared to their Kerr counterparts, but the difference is small compared to the level of variability. We find $\langle \phi_{\rm B,KL} \rangle - \langle \phi_{\rm B,KHL} \rangle = 1.816$, $\langle \phi_{\rm B,KM} \rangle - \langle \phi_{\rm B,KHM} \rangle = 4.543$, while $\sigma_{\rm KL} = 12.684$, $\sigma_{\rm KHL} = 8.106$, $\sigma_{\rm KM} = 16.830$, $\sigma_{\rm KHM} = 12.808$. As mentioned, differences of means are consistently lower than the standard deviations; $\phi_{B,KH}$ is consistent with $\phi_{B,K}$ up to $1\sigma$, even though it looks a little lower. This is consistent with recent work [41], which shows that nonspin gravitational degrees of freedom that reduce the black hole size also suppress the magnetic flux, scaling with the horizon area.

The two bottom panels in Fig. 1 depict the jet efficiency, calculated as

$$\eta = \frac{\dot{M} - \dot{E}}{\langle \dot{M} \rangle}. \tag{10}$$

This definition, consistent with Tchekhovskoy *et al.* [65], does not attempt to define a distinct jet region and thus includes all outflowing energy, regardless of origin. With solid lines, we have overplotted the values expected for a Blandford-Znajek (BZ) jet efficiency from [65] (Eq. 2), where we use the values of $\phi$ from Eq. (9).

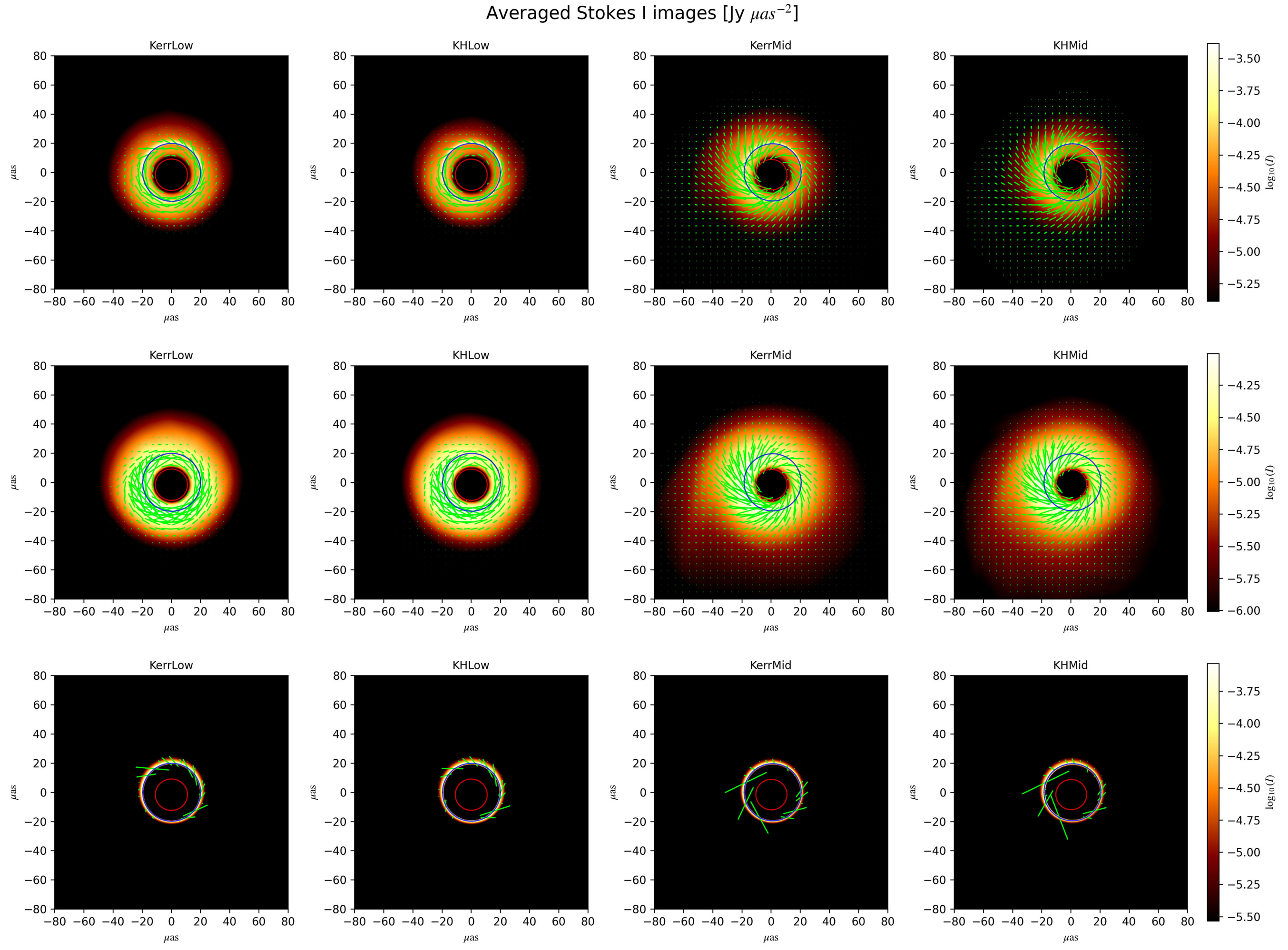

FIG. 2. Time-averaged Stokes $I$ images for the four fiducial models, with polarization ticks overlaid, for $n = \text{all}$ (top), $n = 0$ (middle), and $n = 1$ (bottom), at resolution $400 \times 400$. The columns show, from left to right, KerrLow, KHLow, KerrMid, and KHMid. Colors indicate $\log_{10}(I/\text{Jy } \mu\text{as}^{-2})$. The blue curve denotes the theoretical critical curve and the red curve the theoretical inner shadow. The critical curve provides a close visual match to the bright ringlike emission in all cases, whereas the inner shadow traces the central intensity depression. Visual differences between the Kerr and modified Kerr-Hayward images are small, especially for corresponding spin values.

Directly from the metric, we calculate the horizon radius, calculated by solving $\Delta(r) = 0$ for the largest positive root, and the horizon angular velocity for the modified Kerr-Hayward to be (1.97877, 0.0254743) and (1.8429, 0.137126) for the low- and mid-spin models, respectively. In contrast to the Kerr values, (1.9949, 0.0251) for KerrLow and (1.8660, 0.1339) for KerrMid, the values for the mid-spin models are larger by orders of magnitude ($\sim$0.06 to $\sim$0.8 mean values) with respect to the low-spin models, consistent with the BZ-like scaling (higher efficiency at higher spin). Modified Kerr-Hayward values are slightly higher than Kerr values for the low-spin models and the opposite is true for the mid-spin models, but the differences are only a few percent compared to the general variability in the time series. For the low-spin models, the larger jet efficiency relative to [65] can be attributed to the inclusive definition of outflow power adopted in Eq. (10).

In summary, our Kerr and modified Kerr-Hayward simulations do not exhibit significant differences in terms of accretion rate, jet efficiency, or magnetic flux.

### B. Imaging and polarization domain

In Fig. 2, we plot time-averaged images of all four models. Total intensity (Stokes $I$) is plotted in log scale with two decades of dynamic range. We overplot the linear polarization structure using green ticks, where we tick length scales with the total amount of polarization, and the angle represents the electric vector position angle (EVPA) $\chi = 0.5\arctan(U/Q)$. To search for metric-sensitive features, we plot the total image in the top row but individually plot the $n = 0$ and $n = 1$ subimages[7] in subsequent rows. We compare these images with the "critical curve" in blue, the curve to which photon ring images converge as $n \to \infty$ [6], and the "inner shadow" in red, the image of the event horizon's equator that manifests as a flux depression in disk emission geometries [7]. For modified Kerr-Hayward, these curves are calculated analytically from the metric as detailed in Appendix C.

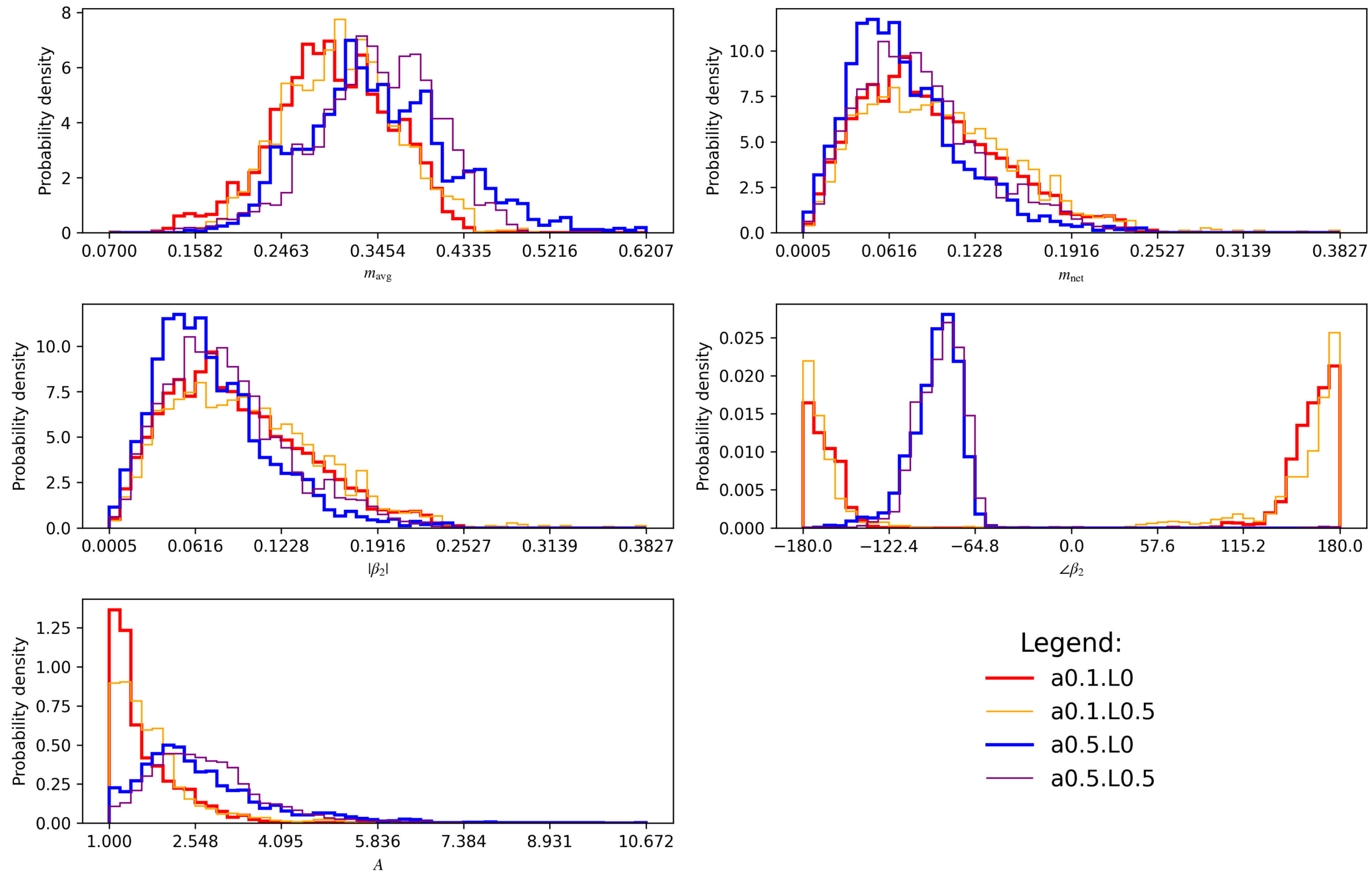

FIG. 3. Distributions of the polarization observables and brightness asymmetry for the four fiducial models, computed from the $n =$ all images at resolution $400 \times 400$, with the blurring kernel 20 μas. The panels show, from top left to bottom left, the average linear polarization $m_{\mathrm{avg}}$, the net linear polarization $m_{\mathrm{net}}$, the quadrupolar polarization amplitude $|\beta_2|$, the quadrupolar polarization phase $\angle\beta_2$, and the brightness asymmetry $A$. All histograms are normalized to unit area. The $m_{\mathrm{avg}}$ and $m_{\mathrm{net}}$ distributions are broadly similar across models, while $|\beta_2|$ and especially $\angle\beta_2$ show stronger spin dependence and more pronounced differences in the high-spin cases. The brightness-asymmetry distributions indicate systematically larger asymmetry in the Kerr models than in the corresponding modified Kerr-Hayward models.

In total intensity, we find that all images faithfully produce a flux depression at their respective inner shadows, which themselves are nearly identical between Kerr and modified Kerr-Hayward models. Because the $n = 0$ image is radially extended, the $n = 1$ image (photon ring) lies just outside the critical curve, as expected. At larger angular scale, the extended emission appears slightly different in the modified Kerr-Hayward model compared to the Kerr model at higher spin, but we attribute these differences to the finite time simulated for the time average.

In linear polarization, the EVPA pattern encodes the magnetic field structure, which in turn is known to encode the BH spin via frame dragging in Kerr simulations [66,67]. We see a more toroidal pattern for $a = 0.1$ and a more radial pattern for $a = 0.5$, as expected from these previous works, because the magnetic field becomes more toroidal as the spin increases. However, the Kerr linear polarization pattern appears very similar to the modified Kerr-Hayward counterpart in each case. This does not seem to be surprising, as the horizon angular velocity is not particularly different between the two considered metrics.

Below, we apply several image metrics to quantify the (lack of) difference between Kerr and modified Kerr-Hayward models in these respects.

### 1. EHT Image Metrics

In Figure 3, we calculate several commonly used image metrics from our IPOLE images defined in Appendix B, with a blurring beam of size 20 μas. Each of these are measurable by the EHT at its current level of resolution. In the first row, we plot the linear polarization fraction, first on spatially resolvable scales of 20 μas ($m_{\mathrm{avg}}$), and next on unresolved scales as may be observed by a single dish ($m_{\mathrm{net}}$). The linear polarization fraction is sensitive to the Faraday rotation depth [which in turn encodes density, temperature, and magnetic field, e.g., [68–70] ] as well as changes to the magnetic field geometry along the line of sight and within the beam. All fiducial models peak at $m_{\mathrm{avg}} \sim 0.3$–$0.35$ and $m_{\mathrm{net}} \sim 0.05$–$0.1$. For ringlike polarization patterns, $m_{\mathrm{net}} < m_{\mathrm{avg}}$ occurs due to symmetry [71]. Kerr and modified Kerr-Hayward models are not distinguishable at this level.

[7]We denote by $n$ the number of half-orbits the photon has orbited the BH before reaching the camera.

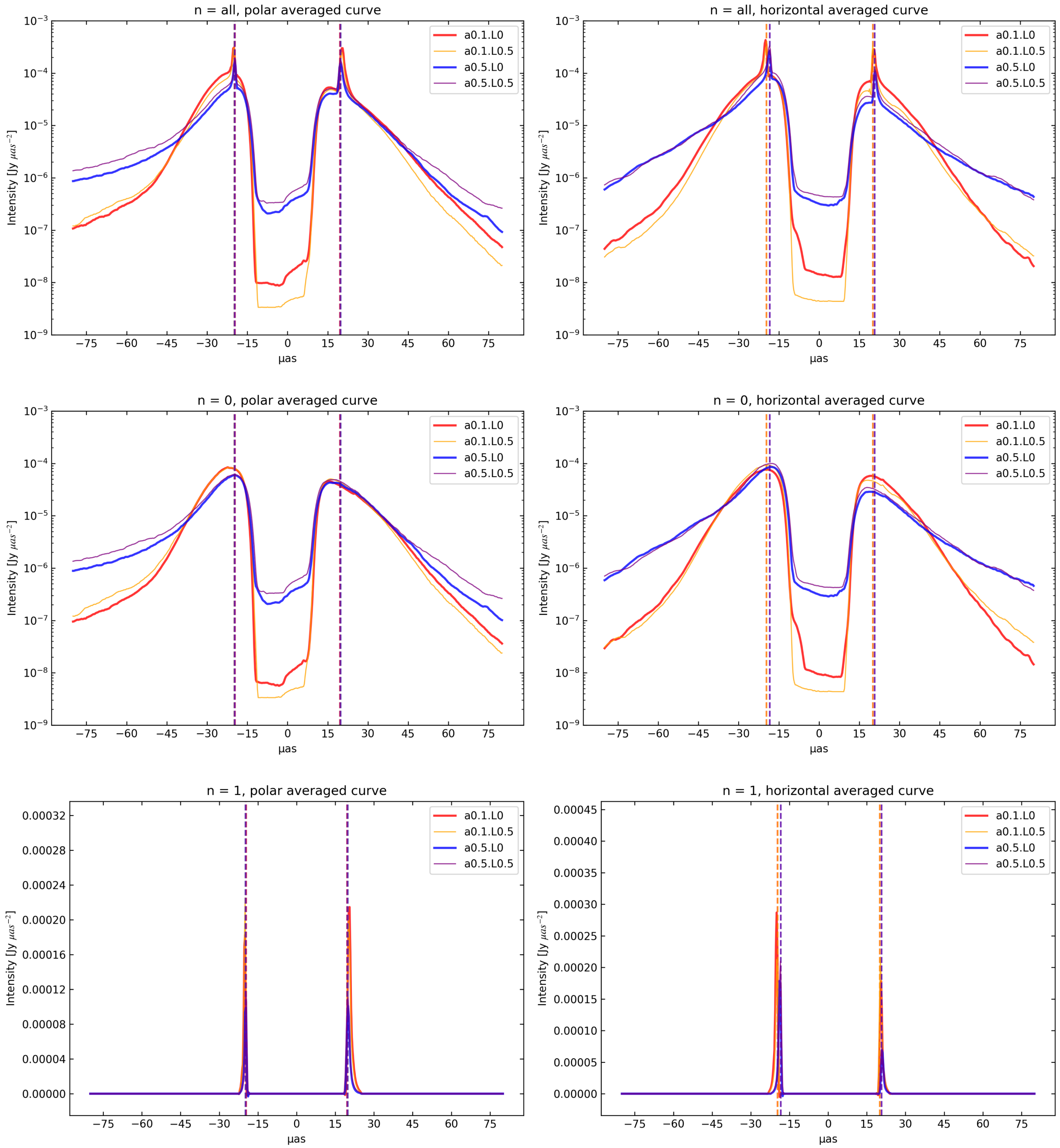

FIG. 4. Time-averaged one-dimensional Stokes $I$ profiles for the four fiducial models, extracted from the image-plane cuts shown separately for $n = \mathrm{all}$ (bottom), $n = 0$ (middle), and $n = 1$ (top), at resolution $400 \times 400$. In each row, the left panel shows the profile along the vertical (polar) cut and the right panel the profile along the horizontal cut. The dashed lines mark the locations of the theoretical critical-curve intercepts in the corresponding direction. These cuts probe the most distant points of the ring structure on the image plane and are used to compare the image-extracted peak locations with the theoretical critical curve. The $n = 1$ profiles are sharply peaked and closely aligned with the critical-curve intercepts, while the $n = 0$ and $n = \mathrm{all}$ profiles are broader and show mild offsets and asymmetries, particularly in the horizontal direction.

TABLE I. Comparison of geometric observables and universal demagnification quantities for the four models. Asymmetry ratios (asym) are listed as (theoretical; image-extracted). $\Delta x_c = |x_{\mathrm{c,image}} - x_{\mathrm{c,theory}}|$ for both the ring and the shadow. For the demagnification scaling tests, $H^{L,R} \equiv \mathrm{HWHM}^{L,R}_{n=1}/\mathrm{HWHM}^{L,R}_{n=0}$ and $e_{L,R} \equiv e^{-\gamma_p^{L,R}}$.

| Model | Asym. (CC) | $\Delta x_c^{\mathrm{ring}}$ | Asym. (IS) | $\Delta x_c^{\mathrm{shadow}}$ | $(R^L, R^R)$ | $F$ | $(H^L, H^R)$ | $(e^{-\gamma_p,L}, e^{-\gamma_p,R})$ |
|---|---|---|---|---|---|---|---|---|
| KerrLow | 0.99997; 0.99830 | 0.02200 | 1.02140; 1.00000 | 0.08500 | (0.37680, 0.25442) | 0.23861 | (0.23098, 0.23839) | (0.04339, 0.04338) |
| KerrMid | 0.99857; 0.99774 | 0.01200 | 1.02077; 1.00000 | 0.86100 | (0.22551, 0.09848) | 0.11772 | (0.26365, 0.21387) | (0.04887, 0.04757) |
| KHLow | 0.99995; 1.00062 | 0.05000 | 1.02141; 1.00000 | 0.23000 | (0.21934, 0.13728) | 0.23446 | (0.23268, 0.23019) | (0.04446, 0.04436) |
| KHMid | 0.99849; 1.00190 | 0.02900 | 1.02074; 1.00861 | 0.85900 | (0.17722, 0.08192) | 0.12514 | (0.25720, 0.22194) | (0.05085, 0.04868) |

The second row plots the amplitude and phase of $\beta_2$, the rotationally symmetric mode of the polarization pattern following a radially averaged Fourier decomposition. The amplitude is correlated with $m_{\mathrm{avg}}$ and encodes the strength of this mode, while the phase encodes the handedness and pitch angle of the EVPA pattern. For $|\beta_2|$, low-spin Kerr peaks at about 0.05 and the rest at 0.075, with high-spin models being narrower and more sharply peaked, and low-spin models having slightly longer tails. The ranges are similar for both Kerr and KH, and we do not see a significant difference in the magnitude. For $\angle\beta_2$, we see a narrow peak around $-90°$ for higher-spin models and a distribution centered around $\pm180°$ for lower-spin models. Compared to [66,67], the distribution of $\angle\beta_2$ for $a = 0.1$ is more symmetric, about $-100°$. This encodes the magnetic field structure around the BH, which emerges through frame dragging [67]. The remarkable agreement between the Kerr and modified Kerr-Hayward models indicates that the dynamics of their respective accretion flows are very similar. Finally, we plot another dynamics-sensitive quantity in the bottom plot, the brightness asymmetry in total intensity as defined by Medeiros *et al.* [72] (Eq. 2), which probes the degree of relativistic Doppler beaming. As expected, larger values of $A$ are achieved as $a$ increases, but we do not see significant changes as $L$ increases.

### 2. Ring Metrics

When considering the "photon ring," we will talk about the following:

(i) *Definition*: *Theoretical ring metrics.* Ring metrics constructed from the spacetime metric, like the Kerr metric or the modified Kerr-Hayward metric (e.g. the critical curve being the feature of the spacetime, expressed in terms of the metrics components) are considered a part of the theoretical ring metrics family.

(ii) *Definition*: *Image ring metrics.* Ring metrics extracted from the image itself. We will talk about the photon ring as the feature of the image, defined in terms of the Stokes I intensity.

In the ideal case, the critical curve and the photon ring expressed in terms of the Stokes I intensity coincide perfectly, and we measure in this section. We define all metrics, both theoretical and image, in Appendix C.

We average all horizontal ($x = 0$) and vertical ($y = 0$) cuts through the intensity plots from all simulation snapshots and plot them in Fig. 4, separately for $n =$ all, 0, 1 (the intensity plots are of the photon rings and the $n$ is the subring index). We choose to take only the horizontal and polar cuts, because those sample the most distant points, and given the difficulty of measuring the ellipticity of the photon ring, we deem that a reasonable measure. On top of the curves we overplot the theoretical critical curves.

In Table I, we present the results of our study of the ring metrics. We compare the asymmetry ratio and the horizontal center displacement $\Delta x_c$, where $\Delta x_c = |x_{\mathrm{c,image}} - x_{\mathrm{c,theory}}|$ of the critical curve versus the $n = 1$ photon ring (image extracted), and see whether the universal demagnification test holds (for $n = 0, 1$), considering three different criteria: the critical curve offset, the subring flux ratio, and the half width half maximum of the intensity ratio (all defined in Appendix C).

From the asymmetry ratio, we see that the critical curve and the extracted photon ring are essentially circular. Based on the horizontal center displacement, we see that the ring center is not much displaced from the center of the critical curve, indicating good overall alignment. Similarly for the inner shadow, the horizontal displacement is small, but it is larger relative to the ring displacement; the overall shape of the shadow remains nearly circular.

Using the described methods of extraction, we find that the peak-extracted offsets for $n = 0, 1$ images produce systematically much larger $R$, $F$ and $HWHM$ ratios than the universal prediction $e^{-\gamma_p}$. This indicates that the lowest-order features do not obey the demagnification scaling, under our extraction, which is consistent with the literature.

## IV. DISCUSSION AND CONCLUSIONS

We compared fiducial Kerr and modified Kerr-Hayward models in the fluid and imaging domains. In the fluid domain, we notice that the Eddington ratio shows strong variability in all models, with low-spin models reaching higher values than mid-spin ones, and mild differences between Kerr and modified Kerr-Hayward. The dimensionless magnetic flux is also comparable across all four models, with Kerr exhibiting slightly higher mean values than modified Kerr-Hayward, which could mean lower magnetic flux accumulation at the horizon in modified Kerr-Hayward. The jet efficiencies increase strongly with spin, which is consistent with Blandford-Znajek predictions. Differences between Kerr and modified Kerr-Hayward are small compared to the temporal variability. Overall, those fluid domain diagnostics indicate that, for the parameters considered here, modified Kerr-Hayward accretion is not distinguishable from Kerr beyond the few-percent level, which is not observable.

In the imaging domain, the polarized intensity images and the EVPA morphology are visually very similar between Kerr and modified Kerr-Hayward for corresponding spins. The critical curve and inner shadow curve (theoretical) match the photon ring and the inner shadow on the images by eye. We do not identify any obvious change in the size or shape of the $n = 1$ subring images between the two spacetimes. We perform more detailed analysis, summarized in the next paragraph. The polarization observables show broadly similar distributions between Kerr and modified Kerr-Hayward. The average polarization peaks at ∼0.3–0.35, while the net polarization stays low, ∼0.05–0.1, which is consistent with large cancellation across ring images. The $\beta_2$ distributions are similar between the metrics, with the peaks being narrower in the mid-spin models, which indicates more ordered EVPA structure at higher spin. For $\arg(\beta_2)$, the mid-spin models show narrow peaks near −90°, while low-spin models have a more degenerate structure. The brightness asymmetry is systematically larger in Kerr than in modified Kerr-Hayward for both spins, which suggests that Doppler-beaming asymmetry could be one of the more promising observables for distinguishing between the two spacetimes.

In the ring metric analysis, the asymmetry ratio indicates that the theoretical critical curve and the extracted photon ring are essentially circular. The horizontal center displacement between the image photon ring and the theoretical critical curve is small in all cases, which indicates a good alignment between the spacetime-defined and the image-defined ring structures. Similarly for the inner shadow, the extracted contour is nearly circular, but the horizontal center displacement is larger than in the photon-ring case. The universal demagnification test does not hold for the extracted $n = 0$ and $n = 1$ features under our current extraction procedure. Measured $R$ and $F$ *HWHM* are systematically larger than the theoretical $e^{-\gamma_p}$, which is consistent with the literature.

For the models that we studied in this paper, the largest differences are driven by spin rather than by the change of the de Sitter Length scale. In the fluid domain, Kerr and modified Kerr-Hayward behave very similarly, all studied parameters differing only at the level of few percent, which is small compared to the intrinsic variability of the simulations themselves. In the image domain, the appearance of the ring and the location of its center remain very similar between the metrics. Among the considered polarization observables, brightness asymmetry and possibly the detailed distribution of $\arg(\beta_2)$ seem to be the most promising quantities for identifying subtle differences between Kerr and modified Kerr-Hayward, but they are very small and might not be well captured observationally. We note the failure of the lowest-order ($n = 0, 1$) subrings to satisfy the universal demagnification scaling under our extraction method, which suggests that these image features are still significantly affected by smoothing and extraction-systematics effects.

Therefore, within the scope of the analysis in this study, modified Kerr-Hayward is observationally very difficult to distinguish from Kerr, both in the plasma dynamics (the fluid domain) and in the polarization structure and ring morphology.

The modified Kerr-Hayward metric considered here eliminates the curvature singularity that characterizes the Kerr geometry. Nevertheless, several important issues remain. In particular, the geodesic completeness of the spacetime must be established, the causality-violating regions require resolution, and the inner horizon must be stabilized against the mass inflation instability. We do not investigate the QNM or linear stability, nor do we address the question of mechanisms of formation of such BHs. Despite these limitations, we expect that the quantum-gravitational effects implicitly modeled by these phenomenological constructions primarily modify the deep interior of the spacetime. Consequently, deviations from the classical geometry near the horizon are likely to be strongly suppressed, and the picture developed here should remain broadly representative of the observable exterior spacetime. Given that, if we assume that certain properties of the Kerr metric are also true for the modified Kerr-Hayward metric, then we find that we cannot distinguish between the two in the EHT pipeline.

We follow the PATOKA pipeline, which consists of running the GRMHD simulations, in which we do not consider the interaction of the spacetime with the inflowing accreting plasma. Our reasoning for this choice is that, considering the ADM mass of the BH, the mass entering it in any reasonable time window (e.g., one EHT observing run or a million years) is minuscule. Due to the type of accretion these objects are undergoing and their extremely low mass accretion rate; see, e.g., Sec. 1 of [73], Sec. 2 of [2]. Moreover, the truncation error in the full numerical relativity simulation would far outweigh the change in mass or self-gravitation. Therefore we view the test fluid approximation to be a reasonable choice, for which GRMHD is an appropriate choice. In the Kerr-Hayward spacetime (KHMid), the energy density of the self-gravitating matter decreases steeply with radius, falling from $\sim 10^{-3}$ to $\sim 10^{-22} \mathrm{M}^{-2}$ over the range $r \sim 1\text{–}10^3 M$. By comparison, the GRMHD plasma energy density is approximately $10^{-18} - 10^{-17} \mathrm{M}^{-2}$, placing it well below the matter energy density in the inner region but comparable at larger radii. We also assume that there is no nongravitational interaction between the inflowing plasma and the background self-gravitating exotic matter that creates the Kerr-Hayward spacetime. We must note, however, that modified Kerr-Hayward is not a vacuum metric (neither on the interior, nor on the exterior), we are therefore assuming that the Kerr-Hayward matter is not interacting with the accreting matter in a meaningful manner. If we discovered that it does, we would require an Einstein solver to introduce a dynamical spacetime evolution.

It is also important to note that, while widely adopted for EHT analysis, GRMHD and GRRT omit important physics needed for fully self-consistent modeling of M87*. The most relevant limitation here is the neglect of radiative cooling: because M87* is thought to be modestly radiatively efficient (about $\sim 10\%$), some dissipated energy remains in the simulations as unphysical heat. Modeling cooling consistently would also require a more satisfactory treatment of electron heating, electron–ion coupling, and dissipation. A further omission is kinetic plasma physics. Since the flow is effectively collisionless, fluid models do not capture effects such as pressure anisotropy and field-aligned heat transport.

The purpose of this work is to determine whether such techniques, limited as they are, would nevertheless be capable of distinguishing between two similar classes of black hole metrics. We find that they would not, even in principle, thus motivating the development of more sophisticated, self-consistent, and physically motivated tooling. This is especially important given future missions like ngEHT and BHEX, that will enable us to resolve the photon ring and perform more rigorous spacetime tests.

## ACKNOWLEDGMENTS

The authors would like to thank Jonathan Gorard for helpful comments and useful conversations. N. B. acknowledges the hospitality of the Center for Computational Astrophysics-Flatiron Institute. N. B. also acknowledges support from the URI Institute for AI & Computational Research. The computations were performed on the UMass-URI UNITY high-performance computing (HPC) cluster hosted at the Massachusetts Green HPC Center. This publication is funded in part by the Gordon and Betty Moore Foundation, Grant No. GBMF-12987. We acknowledge financial support from the National Science Foundation (No. AST-2307887). This work was supported in part by the Black Hole Initiative, which is funded by grants from the John Templeton Foundation (Grant No. 62286) and the Gordon and Betty Moore Foundation (Grant No. GBMF-8273).

However, the opinions expressed in this work are those of the author(s) and do not necessarily reflect the views of these Foundations.

## DATA AVAILABILITY

The code used to analyze the data is available from Ref. [50]. The data that support the findings of this study were generated by numerical simulations. The source code and parameters used to generate the simulations is publicly available [50].

## APPENDIX A: INNER SHADOWS OF MODIFIED KERR-HAYWARD BLACK HOLES

The example implementation of the definitions below, used to generate the results in this study, has been made open-source [50].

Here we present the morphology of inner shadows in modified Kerr-Hayward black holes (BHs). While the main text focuses on two Kerr and two modified Kerr-Hayward BHs, we extend the analysis here to span the full BH parameter space, thereby quantifying the maximal deviations that can arise. In the main text, we restrict attention to nearly polar observers, motivated by our interest in modeling M87* observations. Here, we additionally consider nearly equatorial observers, providing a complementary perspective relevant to observations of Sgr A*, whose inclination remains uncertain; see, however, Fig. 9 of Ref. [74].

The inner shadow is the direct image of the equatorial horizon [7]. We compute it for modified Kerr-Hayward black holes using the integral formulation of the geodesic equations (see, e.g., Refs. [75,76]), solving the associated boundary-value problem via root-finding rather than by ray tracing photons from the image plane.

We consider photon trajectories that originate on the equatorial horizon and reach a distant observer. The emission points lie on the equatorial horizon, parameterized by the azimuthal angle $\varphi_{\mathrm{e}}$, with $r_{\mathrm{e}} = r_+$, $\vartheta_{\mathrm{e}} = \pi/2$, and $0 \leq \varphi_{\mathrm{e}} < 2\pi$, where $r_+$ is the event horizon radius. The observer is located at $r_{\mathrm{o}} = \infty$, $\vartheta_{\mathrm{o}} = i$, and $\varphi = \varphi_{\mathrm{o}}$.

For each emitter location $\varphi_{\mathrm{e}}$, we seek the photon impact parameters $(\eta, \xi)$ such that a null geodesic connects the emission point to the observer. This defines a two-point boundary-value problem: the endpoints of the trajectory are fixed, and the corresponding constants of motion $(\eta_{\rm IS}(\varphi_{\rm e}), \xi_{\rm IS}(\varphi_{\rm e}))$ must be determined. Solving this problem yields the map from the equatorial horizon to the observer's image plane.

The (Boyer-Lindquist) coordinate 4-velocity of a photon is given by

$$\dot{x}^\mu = [\mathcal{T}_r + a^2\cos^2\vartheta, \pm_r\sqrt{\mathcal{R}}, \pm_\vartheta\sqrt{\Theta}, \Phi_r + \xi\csc^2\vartheta], \quad \text{(A1)}$$

where the overdot denotes differentiation with respect to the Mino time $\lambda_{\rm m}$ [77], a nonaffine parameter along the photon orbit. The relevant functions are [45]

$$\mathcal{T}_r = \frac{(r^2+a^2)^2 - 2Fa\xi}{\Delta} - a^2, \quad \text{(A2)}$$

$$\Phi_r = \frac{a(2F - a\xi)}{\Delta}, \quad \text{(A3)}$$

$$\mathcal{R} = [(r^2+a^2) - a\xi]^2 - \Delta\eta^2, \quad \text{(A4)}$$

$$\Theta = \eta + (\xi - a)^2 - (\xi\csc\vartheta - a\sin\vartheta)^2, \quad \text{(A5)}$$

where $\Delta$ and $F$ are defined in Eq. (3).

Photons emitted from the equatorial horizon and received by the observer have strictly positive radial velocity, i.e., they exhibit no radial turning points ($\dot{r} = 0 \Leftrightarrow \mathcal{R} = 0$). Two qualitatively distinct classes of direct photon trajectories contribute to the inner shadow[8]: those without polar turning points ($n_{\rm ptp} = 0$) and those with a single turning point ($n_{\rm ptp} = 1$). We direct the reader to Fig. 14 of Ref. [78] for intuition. A polar turning point is defined by $\dot{\vartheta} = 0 \Leftrightarrow \Theta = 0$.

Without loss of generality, we assume that the observer is in the northern hemisphere ($0 \le i \le \pi/2$) and that $\varphi_{\rm o} = 0$. For an observer in the southern hemisphere at $\vartheta_{\rm o} = \pi - i$, the inner shadow is the mirror image of that seen at inclination $i$, reflected about the image-plane $y$-axis (the Bardeen $\beta$-axis). With these conventions, the boundary-value problem reduces to the following system:

*Case* $n_{\rm ptp} = 0$:

$$\int_{r_+}^{\infty} \frac{{\rm d}r}{\sqrt{\mathcal{R}}} = -\int_{\pi/2}^{i} \frac{{\rm d}\vartheta}{\sqrt{\Theta}}, \quad \text{(A6a)}$$

$$\varphi_{\rm o} - \varphi_{\rm e} = \left[\int_{r_+}^{\infty} \frac{\Phi_r {\rm d}r}{\sqrt{\mathcal{R}}} - \int_{\pi/2}^{i} \frac{\xi\csc^2\vartheta{\rm d}\vartheta}{\sqrt{\Theta}}\right] \bmod 2\pi, \quad \text{(A6b)}$$

[8]As an aside, we note that all photon orbits contributing to the inner shadow are necessarily "ordinary" null geodesics, in the sense of Ref. [75], since they must reach the equatorial plane.

*Case* $n_{\rm ptp} = 1$:

$$\int_{r_+}^{\infty} \frac{{\rm d}r}{\sqrt{\mathcal{R}}} = -\int_{\pi/2}^{\vartheta_-} \frac{{\rm d}\vartheta}{\sqrt{\Theta(\vartheta)}} + \int_{\vartheta_-}^{i} \frac{{\rm d}\vartheta}{\sqrt{\Theta(\vartheta)}}, \quad \text{(A6c)}$$

$$\varphi_{\rm o} - \varphi_{\rm e} = \left[\int_{r_+}^{\infty} \frac{\Phi_r {\rm d}r}{\sqrt{\mathcal{R}}} - \int_{\pi/2}^{\vartheta_-} \frac{\xi\csc^2\vartheta{\rm d}\vartheta}{\sqrt{\Theta}} + \int_{\vartheta_-}^{i} \frac{\xi\csc^2\vartheta{\rm d}\vartheta}{\sqrt{\Theta}}\right] \bmod 2\pi, \quad \text{(A6d)}$$

where $\vartheta_-$ is the polar turning point (i.e., $\Theta(\vartheta_-) = 0$) in the northern hemisphere,

$$\vartheta_- = \arccos\left[\sqrt{\Delta_\vartheta + \sqrt{\Delta_\vartheta^2 + \frac{\eta}{a^2}}}\right], \quad \text{(A7)}$$

$$\Delta_\vartheta = \frac{1}{2}\left[1 - \frac{\eta + \xi^2}{a^2}\right]. \quad \text{(A8)}$$

We solve the coupled integral equations (A6) for $(\eta_{\rm IS}, \xi_{\rm IS})$ at each emitter location $\varphi_{\rm e}$. At each $\varphi_{\rm e}$, a solution is obtained either for $n_{\rm ptp} = 0$ ("near side" in $\varphi$) or for $n_{\rm ptp} = 1$ (far side).

The corresponding image-plane coordinates are obtained via the standard transformation to Bardeen coordinates,

$$\alpha_{\rm IS}(\varphi_{\rm e}) = -\xi_{\rm IS}(\varphi_{\rm e})\csc i, \quad \text{(A9)}$$

$$\beta_{\rm IS}(\varphi_{\rm e}) = \pm\sqrt{\Theta(i, \eta(\varphi_{\rm e}), \xi(\varphi_{\rm e}))}. \quad \text{(A10)}$$

We choose the negative sign for $n_{\rm ptp} = 0$ and the positive sign for $n_{\rm ptp} = 1$, in accordance with the sign of the photon's polar velocity at the observer: photons that reach the observer without a polar turning point have negative polar velocity, whereas those that do encounter a turning point arrive with positive polar velocity.

Our code computes the inner shadow for arbitrary Johannsen [79] BH spacetimes. For a discussion of photon rings and the associated critical parameters in this class of spacetimes, see Refs. [76,80].

Figure 5 shows the shadow and inner-shadow curves for a representative set of modified Kerr-Hayward black holes with fixed ADM mass $M$. The parameter space is sampled uniformly in spin $a$ and de Sitter length scale $L$, restricted to the subextremal region.

The dependence on $L$, which controls the magnitude of deviations from the Kerr geometry, is weak for both the shadow and the inner shadow across all viewing angles. For near-polar inclinations, both curves remain nearly circular, reflecting the approximate symmetry of the spacetime about the rotation axis.

At higher inclinations, however, their behavior diverges. The shadow remains comparatively close to circular,

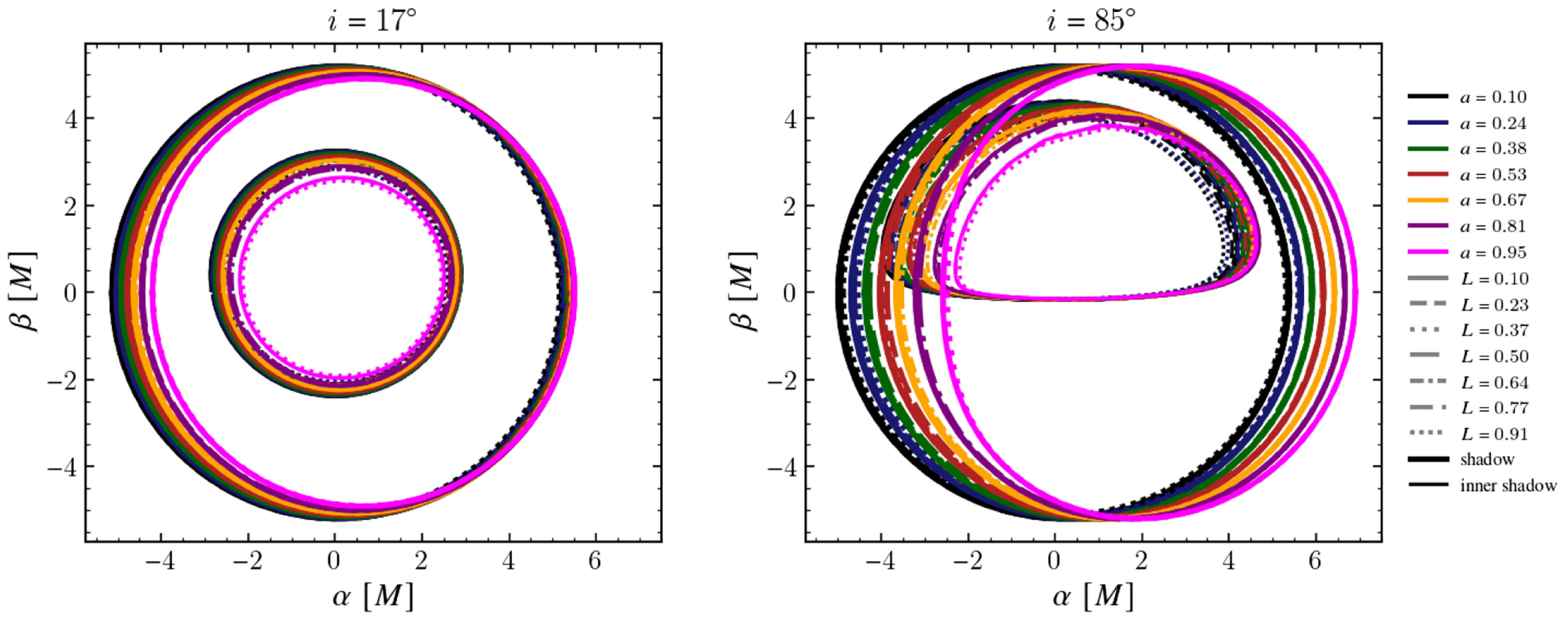

FIG. 5. Shadows and inner shadows of modified Kerr-Hayward black holes for nearly polar (left) and nearly equatorial (right) observers. For polar viewing angles, both structures are nearly circular, with weak dependence on the spin $a$ and the de Sitter length scale $L$. For equatorial observers, the shadow remains approximately circular but is increasingly displaced to the right with increasing spin due to frame dragging, while its dependence on $L$ is subdominant. In contrast, the inner shadow is strongly noncircular: its flattened segment corresponds to the near side of the equatorial horizon, while the upper portion traces the far side. The close approach of the inner shadow to the shadow boundary arises from photons emitted from behind the black hole that undergo deflections close to $\pi$, corresponding to nearly first-order (indirect) image trajectories.

exhibiting only a gradual increase in asymmetry and a modest displacement with increasing spin due to frame dragging. In contrast, the inner shadow develops pronounced noncircularity. Its flattened segment corresponds to emission from the near side of the equatorial horizon, while the upper portion traces photons originating from the far side. The close approach of the inner shadow to the shadow boundary arises from photons that undergo large deflections, approaching $\pi$, corresponding to nearly first-order (indirect) image trajectories.

These geometric differences between the shadow and inner shadow, particularly their distinct responses to inclination and spin, motivate the quantitative analysis of their sizes, shapes, and displacements presented in Fig. 6.

Figure 6 provides a quantitative characterization of the shadow and inner-shadow properties as functions of spin, de Sitter length scale, and observer inclination. To facilitate comparison with previous work, we adopt the standard moment-based definitions of the mean radius $\bar{r}$ and eccentricity $e$ from Appendix B of Ref. [7], applied to closed convex curves enclosing the image-plane origin (such as those shown in Fig. 5). The centroid offset is defined as the magnitude of the first-moment vector, $\Delta C = \sqrt{\mu_\alpha^2 + \mu_\beta^2}$.

The behavior of the radii reflects the underlying structure of the photon shell. The mean shadow radius (panel a) decreases with decreasing inclination because larger-radius spherical photon orbits, particularly those near the equatorial plane, do not contribute to the observed boundary at low inclinations. Its decrease with increasing spin or de Sitter length scale reflects a reduction in the horizon area and the associated contraction of the photon shell. Overall, the mean shadow radius varies only weakly across the parameter space, with a maximum change of ≈6% relative to the Schwarzschild value.

The mean inner-shadow radius (panel d) exhibits a similar but more pronounced trend. For nearly polar observers, the contributing photons undergo a total polar deflection of $\approx\pi/2$, whereas for nearly equatorial observers, far-side emission is imaged by photons undergoing large deflections ($\approx\pi$), requiring larger impact parameters and producing a larger inner shadow. The decrease in the inner-shadow size with increasing spin and de Sitter length scale likewise reflects a reduction in the horizon size. In this case, the variation is significantly larger, reaching ≈30% relative to the Schwarzschild equatorial value.

The eccentricities quantify the shape distortions visible in Fig. 5. The shadow eccentricity (panel b) is most sensitive to spin $a$, with a smaller dependence on inclination $i$ and a weak dependence on the de Sitter length scale $L$. In contrast, the inner-shadow eccentricity (panels e1, e2) is dominated by inclination, with only a very weak dependence on spin and negligible dependence on $L$.

The centroid offsets capture the displacement of the image due primarily to frame dragging. The shadow centroid offset (panel c) is most sensitive to spin, with a smaller contribution from inclination and a weak dependence on $L$. The inner-shadow centroid offset (panels f1, f2) is instead most sensitive to inclination, followed by $L$, with a weaker dependence on spin.

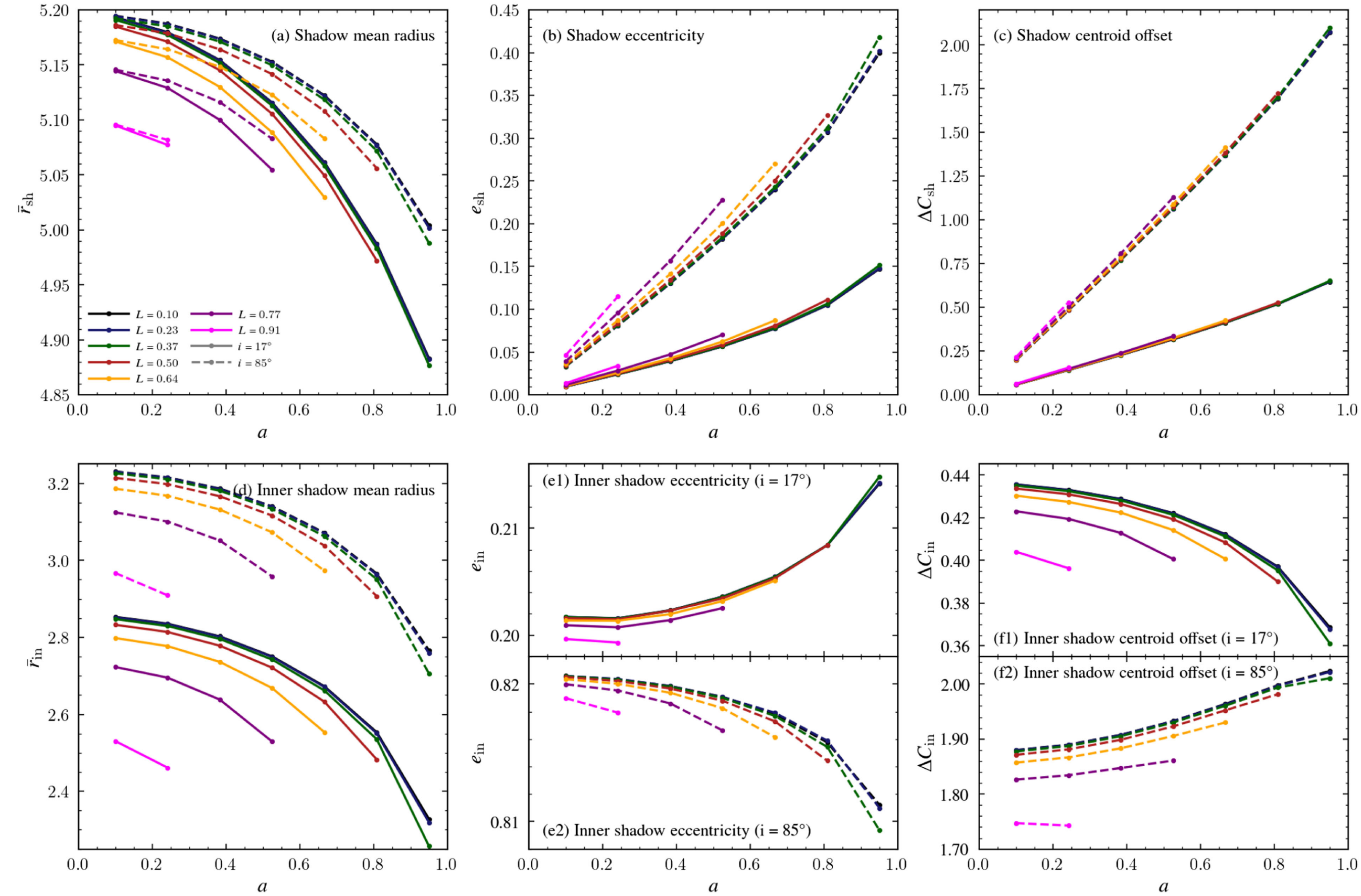

FIG. 6. Radius, eccentricity, and centroid offset of the shadow (top) and inner shadow (bottom). Among these observables, the shadow eccentricity and centroid offset are the most effective probes of the spin $a$, while the inner-shadow eccentricity and centroid offset are most sensitive to the inclination $i$. The de Sitter length scale $L$ imprints comparatively weaker signatures across all characteristics, with its effects being most pronounced in the inner-shadow radius and centroid offset. These results indicate that a joint analysis of shadow and inner-shadow observables is necessary to constrain the underlying parameters.

Taken together, these results show that shadow observables primarily constrain the spin, while inner-shadow observables provide stronger sensitivity to the inclination, with the de Sitter length scale producing comparatively weaker signatures across all metrics. A joint analysis of shadow and inner-shadow characteristics is therefore required to disentangle the underlying parameters.

## APPENDIX B: ADDITIONAL DEFINITIONS

We introduce several image and polarization domain metrics used throughout this study.

Following [66], we define the average linear polarization as

$$m_{\rm avg} \equiv \frac{\sum_{\rm pix}\sqrt{Q^2+U^2}}{\sum_{\rm pix} I}, \tag{B1}$$

and the net linear polarization fraction as

$$m_{\rm net} \equiv \frac{\sqrt{(\sum_{\rm pix} Q)^2 + (\sum_{\rm pix} U)^2}}{\sum_{\rm pix} I}, \tag{B2}$$

where the sums are taken over image pixels.

We also investigate the polarimetric rotational modes. Defining the complex linear polarization on the image plane by

$$P(\alpha,\beta) \equiv Q(\alpha,\beta) + iU(\alpha,\beta), \tag{B3}$$

the complex mode coefficients are

$$\beta_m \equiv \frac{\int_{\mathcal{A}} P(\alpha,\beta) e^{-im\varphi} dA}{\int_{\mathcal{A}} I(\alpha,\beta) dA}, \tag{B4}$$

where $\varphi$ is the image plane polar angle and $\mathcal{A}$ denotes the image region (or annulus) over which the mode decomposition is performed. We are particularly interested in the quadrupolar mode amplitude,

$$|\beta_2| \equiv \sqrt{\Re(\beta_2)^2 + \Im(\beta_2)^2}, \tag{B5}$$

and the corresponding phase,

$$\arg(\beta_2) \equiv \mathrm{angle}(\beta_2). \tag{B6}$$

Additionally, following Medeiros *et al.* [[72] their Eq. 2], we define the brightness asymmetry $A$, which measures relativistic Doppler-beaming asymmetry, as

$$A = \frac{\int_{2\mathrm{asin}(i)}^{r_{\mathrm{out}}} I(X, Y=0)dX}{\int_{-r_{\mathrm{out}}}^{2\mathrm{asin}(i)} I(X, Y=0)dX}, \tag{B7}$$

where $I(X, Y=0)$ is the image brightness across the cross section and $r_{\mathrm{out}}$ is the outer radius of simulated images.

## APPENDIX C: ANALYTIC PHOTON RING AND INNER SHADOW CALCULATIONS

**Theoretical ring metrics:**

(i) Inner shadow

For the nearly polar observers considered in this study (e.g., $i = 163°$, equivalently $17°$ up to the $\beta$-axis reflection discussed in Appendix A), the inner shadow is obtained from the same Bardeen map,

$$\alpha_{\mathrm{IS}} = -\xi_{\mathrm{IS}} \csc i, \quad \beta_{\mathrm{IS}} = \pm\sqrt{\Theta(i;\eta_{\mathrm{IS}}, \xi_{\mathrm{IS}})}, \tag{C1}$$

with

$$\Theta(i;\eta,\xi) = \eta + (\xi - a)^2 - (\xi \csc\, i - a \sin\, i)^2. \tag{C2}$$

Therefore, for both Kerr and modified Kerr-Hayward spacetimes, the distinction does not lie in the image-plane transformation itself, but in the values of $(\xi_{\mathrm{IS}}, \eta_{\mathrm{IS}})$, which are obtained by solving the corresponding null-geodesic boundary-value problem in the chosen metric.

(ii) Critical curve

Here, the critical curve is parameterized by the spherical-null-geodesic radius $r$, with image-plane coordinates $(\alpha, \beta)$,

$$\alpha_{\mathrm{SNG}}(r) = -\xi_{\mathrm{SNG}}(r) \csc\, i, \\ \beta_{\mathrm{SNG}}(r) = \pm\sqrt{\Theta_{\mathrm{SNG}}(r, i)}, \tag{C3}$$

where

$$\Theta_{\mathrm{SNG}}(r,i) = \eta^2_{\mathrm{SNG}}(r) - \frac{[\xi_{\mathrm{SNG}}(r) - a \sin^2 i]^2}{\sin^2 i}, \tag{C4}$$

and

$$\xi_{\mathrm{SNG}}(r) = \frac{r^2 + a^2}{a} - \frac{4r\Delta}{a\Delta'}, \tag{C5}$$

$$\eta_{\mathrm{SNG}}(r) = \frac{4r\sqrt{\Delta}}{\Delta'}. \tag{C6}$$

These quantities are obtained by solving the spherical null geodesic conditions $\mathcal{R}(r, \eta, \xi) = 0$ and $\mathcal{R}'(r, \eta, \xi) = 0$, which enforce vanishing radial velocity and acceleration, respectively. The explicit form of $\mathcal{R}$ is given in Eq. (A5). Therefore, for both Kerr and modified Kerr-Hayward black holes, the critical-curve map itself is unchanged, with the difference only in the metric function $\Delta(r)$, and hence in the resulting $\xi_{\mathrm{SNG}}(r)$ and $\eta_{\mathrm{SNG}}(r)$.

(iii) Diameter:

– horizontal:

$$D_\alpha = \max(\alpha) - \min(\alpha), \tag{C7}$$

– vertical:

$$D_\beta = \max(\beta) - \min(\beta). \tag{C8}$$

(iv) Asymmetry ratio:

$$\mathrm{ratio} = \frac{D_\alpha}{D_\beta}. \tag{C9}$$

(v) Horizontal center displacement:

$$x_c = \frac{1}{2}(\alpha_{\max} + \alpha_{\min}). \tag{C10}$$

**Image metrics**

(i) Photon ring

Let $I(\alpha, \beta)$ be the image intensity smoothed with a Savitzky–Golay filter. We define the horizontal and vertical cuts

$$I_\alpha(x) \equiv I(x, 0), \qquad I_\beta(y) \equiv I(0, y). \tag{C11}$$

For a generic cut $I(u)$, let

$$u_{\min} = \arg \min_u I(u), \tag{C12}$$

$$u^L_{\mathrm{peak}} = \arg \max_{u<u_{\min}} I(u), \tag{C13}$$

$$u^R_{\mathrm{peak}} = \arg \max_{u>u_{\min}} I(u), \tag{C14}$$

for the central minimum and left/right global peaks.

For a fixed $\eta \in (0, 1)$, with $\eta = 0.99$ in this work, define the thresholds

$$T_L = I(u_{\min}) + \eta[I(u^L_{\text{peak}}) - I(u_{\min})], \quad \text{(C15)}$$

$$T_R = I(u_{\min}) + \eta[I(u^R_{\text{peak}}) - I(u_{\min})]. \quad \text{(C16)}$$

We define the left/right intercepts as the inward threshold crossings,

$$u_L\colon\ I(u_L) = T_L, \qquad u_L \in [u^L_{\text{peak}}, u_{\min}], \quad \text{(C17)}$$

$$u_R\colon I(u_R) = T_R, \qquad u_R \in [u_{\min}, u^R_{\text{peak}}], \quad \text{(C18)}$$

found numerically by bracketing and linear interpolation, and identify

$$\alpha_{\min} = u_L[I_\alpha], \qquad \alpha_{\max} = u_R[I_\alpha], \quad \text{(C19)}$$

$$\beta_{\min} = u_L[I_\beta], \qquad \beta_{\max} = u_R[I_\beta]. \quad \text{(C20)}$$

(ii) Inner shadow

For a generic cut $I(u)$ defined above, let $u_{\min}$ be the central minimum, in practice within $|u| \le u_0$, and define

$$d(u) \equiv \frac{dI}{du}, \qquad d_{\max} \equiv \max_u |d(u)|. \quad \text{(C21)}$$

For fixed $f \in (0, 1)$ and $p \in \mathbb{N}$, with $f = \texttt{frac}$ and $p = \texttt{persist}$, set the slope threshold

$$\tau = f d_{\max}. \quad \text{(C22)}$$

We define the left/right intercepts as the first outward locations where the slope exceeds threshold for $p$ consecutive samples:

$$u_L = \max\Big\{u < u_{\min}\colon\ d(u), d(u - \Delta u), \ldots, d(u - (p-1)\Delta u) < -\tau\Big\}, \quad \text{(C23)}$$

$$u_R = \min\Big\{u > u_{\min}\colon d(u), d(u + \Delta u), \ldots, d(u + (p-1)\Delta u) > \tau\Big\}, \quad \text{(C24)}$$

found by scanning away from $u_{\min}$, and identify

$$\alpha_{\min} = u_L[I_\alpha], \qquad \alpha_{\max} = u_R[I_\alpha], \quad \text{(C25)}$$

$$\beta_{\min} = u_L[I_\beta], \qquad \beta_{\max} = u_R[I_\beta]. \quad \text{(C26)}$$

(iii) Diameter:

– horizontal:

$$D_x = x_R - x_L. \quad \text{(C27)}$$

– vertical:

$$D_y = y_R - y_L. \quad \text{(C28)}$$

(iv) Asymmetry ratio:

$$\text{ratio} = \frac{D_x}{D_y}. \quad \text{(C29)}$$

(v) Horizontal center displacement:

$$x_c = \frac{1}{2}(x_R + x_L). \quad \text{(C30)}$$

Universal demagnification test. This quantity is tied to how quickly successive subrings pile up toward the critical curve. The higher-order subrings ($n \ge 2$) follow a simple scaling relation parameterized by a demagnification exponent $\gamma_p$ [76,80–82]. We check whether the $n = 0$ and $n = 1$ rings follow a similar behavior, as they may become observable with upcoming space missions.

For each side of the peak, we calculate three types of ratios that all scale for large $n$.

(vi) Critical-curve offset $R^{(s)}$

$$\delta x_0^{(s)} \equiv |x_0^{(s)} - x_{\text{CC}}^{(s)}|, \quad \text{(C31)}$$

$$\delta x_1^{(s)} \equiv |x_1^{(s)} - x_{\text{CC}}^{(s)}|. \quad \text{(C32)}$$

Here, $x_{\text{CC}}^{(s)}$ is the critical-curve intercept and $\delta x_n^{(s)}$ is the outer peak location in the $n = 0, 1$ image for side $s = L, R$. Then,

$$R^{(s)} \equiv \frac{\delta x_1^{(s)}}{\delta x_0^{(s)}}. \quad \text{(C33)}$$

We say that universality holds if

$$R^{(s)} \approx e^{-\gamma_p}. \quad \text{(C34)}$$

(vii) Angle-dependent subring flux ratio
We say that universality holds if [6]

$$\frac{F^{n+1}_{\rm ring}}{F^{n}_{\rm ring}} \approx e^{-\gamma_p}. \quad \text{(C35)}$$

(viii) Half width at half maximum ratio
We say that universality holds if

$$\frac{\rm HWHM^{n+1}}{\rm HWHM^{n}} \approx e^{-\gamma_p}. \quad \text{(C36)}$$

Here, $\gamma_p = \gamma_p(r_{\rm SNG}; a, L)$ is the theoretical demagnification exponent [80]. It measures the radial instability of the unstable spherical photon orbit that generates the same point on the critical curve.